\documentclass[prl,aps,preprint,showpacs]{revtex4}
\usepackage{graphicx}

\begin{document}
\title{Thermal expansion of Josephson junctions as an
elastic response to an effective stress field}
\author{S. Sergeenkov$^{1,2}$, G. Rotoli$^{3}$, G. Filatrella$^{4}$, and F.M. Araujo-Moreira$^{1}$}
\affiliation{$^{1}$Grupo de Materiais e Dispositivos, Centro
Multidisciplinar para o Desenvolvimento de Materiais Ceramicos,
Departamento de F\'{i}sica, Universidade Federal de S\~{a}o
Carlos, Caixa Postal 676 - 13565-905 S\~{a}o Carlos, SP, Brazil\\
$^{2}$Laboratory of Theoretical Physics, Joint
Institute for Nuclear Research, 141980 Dubna, Moscow Region, Russia\\
$^{3}$CNISM and DIMEG, Universit\`{a} di L'Aquila,
Localit\'{a} Monteluco, I-67040 L'Aquila, Italy \\
$^{4}$Laboratorio Regionale SuperMat CNR-INFM Salerno and
Dipartimento di Scienze Biologiche ed Ambientali, Universit\`{a}
del Sannio,  Via Port'Arsa, 11, I-82100 Benevento, Italy}

%\preprint

\date{\today}

\begin{abstract}
By introducing a concept of thermal expansion (TE) of a Josephson
junction as an elastic response to an effective stress field, we
study (both analytically and numerically) the temperature and
magnetic field dependence of TE coefficient $\alpha $ in a single
small junction and in a square array. In particular, we found that
in addition to {\it field} oscillations due to Fraunhofer-like
dependence of the critical current, $\alpha $ of a small single
junction also exhibits strong flux driven {\it temperature}
oscillations near $T_C$. We also numerically simulated stress
induced response of a closed loop with finite self-inductance (a
prototype of an array) and found that $\alpha $ of a $5\times 5$
array may still exhibit temperature oscillations provided the
applied magnetic field is strong enough to compensate for the
screening induced effects.
\end{abstract}


\maketitle

\section{I. Introduction}

Inspired by new possibilities offered by the cutting-edge
nanotechnologies,  the experimental and theoretical physics of
increasingly sophisticated mesoscopic quantum devices (heavily
based on Josephson junctions and their arrays) is becoming one of
the most exciting and rapidly growing areas of modern science
(see, e.g.,~\cite{a1,a2,a3,a4} for the recent reviews on charge
and spin effects in mesoscopic 2D Josephson junctions and
quantum-state engineering with Josephson-junction devices). In
particular, a remarkable increase of the measurements technique
resolution  made it possible to experimentally detect such
interesting phenomena as flux avalanches~\cite{a5} and geometric
quantization~\cite{a6} as well as flux dominated behavior of heat
capacity~\cite{a7} in Josephson junctions (JJs) and their arrays
(JJAs).

At the same time, given a rather specific
magnetostrictive~\cite{a8} and piezomagnetic~\cite{a9} response of
Josephson systems, one can expect some nontrivial behavior of the
thermal expansion (TE) coefficient in JJs as well. Of special
interest are the properties of TE in applied magnetic field. For
example, some superconductors like $Ba_{1-x}K_xBiO_3$,
$BaPb_xBi_{1-x}O_3$ and $La_{2-x}Sr_xCuO_4$ were found~\cite{a10}
to exhibit anomalous temperature behavior of both magnetostriction
and TE which were attributed to the field-induced suppression of
the superstructural ordering in the oxygen sublattices of these
systems.

By introducing a concept of TE of Josephson contact (as an elastic
response of JJ to an effective stress field), in the present paper
we consider the temperature and magnetic field dependence of TE
coefficient $\alpha (T,H)$ in a small single JJ and in a single
plaquette (a prototype of the simplest JJA). In a short contact,
the field-induced $\alpha (T,H)$ is found to exhibit strong
temperature oscillations near $T_C$. At the same time, in an array
(described via a closed loop with finite self-inductance) for
these oscillations to manifest themselves, the applied field
should be strong enough to overcome the screening induced
self-field effects.

\section{II. Thermal expansion of a small Josephson contact}

Since thermal expansion coefficient $\alpha (T,H)$ is usually
measured using mechanical dilatometers~\cite{1}, it is natural to
introduce TE as an elastic response of the Josephson contact to an
effective stress field $\sigma $~\cite{a9,2}. Namely, we define
the TE coefficient (TEC) $\alpha (T,H)$ as follows:
% 1
\begin{equation}
\alpha (T,H)=\frac{d \epsilon }{dT}
 \label{alphadefinition}
\end{equation}
where an appropriate strain field $\epsilon$ in the contact area
is related to the Josephson energy $E_J$ as follows ($V$ is the
volume of the sample):
% 2
\begin{equation}
\epsilon =-\frac{1}{V}\left [\frac{dE_J}{d\sigma}\right ]_{\sigma
=0}
\end{equation}
For simplicity and to avoid self-field effects, we start with a
small Josephson contact of length $w<\lambda _J$ ($\lambda
_J=\sqrt{\Phi _0/\mu _0dj_{c}}$ is the Josephson penetration
depth) placed in a strong enough magnetic field (which is applied
normally to the contact area) such that $H>\Phi _0/2\pi \lambda
_Jd$, where $d=2\lambda _{L}+t$, $\lambda _{L}$ is the London
penetration depth, and $t$ is an insulator thickness.

The Josephson energy of such a contact in applied magnetic field
is governed  by a Fraunhofer-like dependence of the critical
current~\cite{orlando}:
\begin{equation}
E_J=J\left (1-\frac{\sin \varphi}{\varphi}\cos \varphi _0\right ),
 \label{Josenergy}
\end{equation}
where $\varphi =\pi \Phi /\Phi _0$ is the frustration parameter
with $\Phi =Hwd$ being the flux through the contact area, $\varphi
_0$ is the initial phase difference through the contact, and
$J\propto e^{-t/\xi}$ is the zero-field tunneling Josephson energy
with $\xi$ being a characteristic (decaying) length and $t$ the
thickness of the insulating layer. The neglected here self-field
effects (screening) will be treated in the next Section for an
array.

Notice that in non-zero applied magnetic field $H$, there are two
stress-induced contributions to the Josephson energy $E_J$, both
related to decreasing of the insulator thickness under pressure.
Indeed, according to the experimental data~\cite{2}, the tunneling
dominated critical current $I_c$ in granular high-$T_C$
superconductors was found to exponentially increase under
compressive stress, viz. $I_c(\sigma )=I_c(0)e^{\kappa \sigma }$.
More specifically, the critical current at $\sigma =9 kbar$ was
found to be three times higher its value at $\sigma =1.5 kbar$,
clearly indicating a weak-links-mediated origin of the phenomenon.
Hence, for small enough $\sigma $ we can safely assume
that~\cite{a9} $t(\sigma )\simeq t(0)(1-\beta \sigma/\sigma_0)$
with $\sigma _0$ being some characteristic value (the parameter
$\beta $ is related to the so-called ultimate stress $\sigma _m$
as $\beta =\sigma _0/\sigma _m$). As a result, we have the
following two stress-induced effects in Josephson contacts:

(I) amplitude modulation leading to the explicit stress dependence
of the zero-field energy
\begin{equation}
J(T,\sigma )=J(T,0)e^{\gamma \sigma/\sigma_0}
\end{equation}
with $\gamma =\beta t(0)/\xi$, and\\

(II) phase modulation leading to the explicit stress dependence of
the flux
\begin{equation}
\Phi (T,H,\sigma )=Hwd(T,\sigma )
\end{equation}
with
\begin{equation}
d(T,\sigma )=2\lambda _{L}(T)+t(0)(1-\beta \sigma/\sigma_0 )
\end{equation}

Finally, in view of Eqs.(1)-(6), the temperature and field
dependence of the small single junction TEC reads (the initial
phase difference is conveniently fixed at $\varphi _0=\pi$):
\begin{equation}
\alpha (T,H)=\alpha (T,0)\left [1+F(T,H)\right ]+\epsilon
(T,0)\frac{dF(T,H)}{dT}, \label{TEcoefficient}
\end{equation}
where
\begin{equation}
F(T,H)=\left [\frac{\sin \varphi}{\varphi}+\frac{\xi}{d(T,0)}
\left (\frac{\sin \varphi}{\varphi}-\cos \varphi \right )\right ],
\end{equation}
with
\begin{equation}
\varphi(T,H)=\frac{\pi \Phi (T,H,0)}{\Phi _0}=\frac{H}{H_0(T)},
\end{equation}
\begin{equation}
\alpha (T,0)=\frac{d\epsilon (T,0)}{dT},
\end{equation}
and
\begin{equation}
\epsilon (T,0)=-\left (\frac{\Phi _0}{2\pi}\right )\left
(\frac{2\gamma}{V \sigma_0}\right )I_C(T).
\end{equation}
Here, $H_0(T)=\Phi _0/\pi wd(T,0)$ with $d(T,0)=2\lambda
_{L}(T)+t(0)$.
\begin{figure*}
\includegraphics[width=8.5cm]{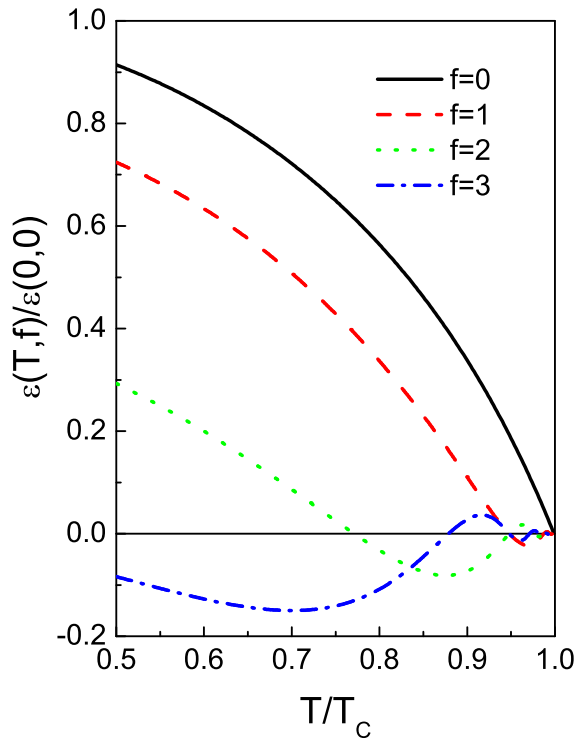}
\caption{\label{fig:1} (Color online) Temperature dependence of
the normalized flux driven strain field $\epsilon (T,f)/\epsilon
(0,0)$ in a single short contact for different values of the
frustration parameter $f=H/H_0(0)$ according to Eqs.(1)-(13).}
\end{figure*}
For the explicit temperature dependence of $J(T,0)=\Phi
_0I_C(T)/2\pi$ we use the well-known~\cite{3} analytical
approximation of the BCS gap parameter (valid for all
temperatures), $\Delta (T)=\Delta (0)\tanh
\left(2.2\sqrt{\frac{T_{C}-T}{T}}\right)$ with $\Delta
(0)=1.76k_BT_C$ which governs the temperature dependence of the
Josephson critical current
\begin{equation}
I_C(T)=I_C(0)\left[ \frac{\Delta (T)}{\Delta (0)}\right] \tanh
\left[ \frac{\Delta (T)}{2k_{B}T}\right]
\end{equation}
while the temperature dependence of the London penetration depth
is governed by the two-fluid model:
\begin{equation}
\lambda _{L}(T)=\frac{\lambda _{L}(0)}{\sqrt{1-(T/T_C)^2}}
\end{equation}

\begin{figure*}
\includegraphics[width=8.5cm]{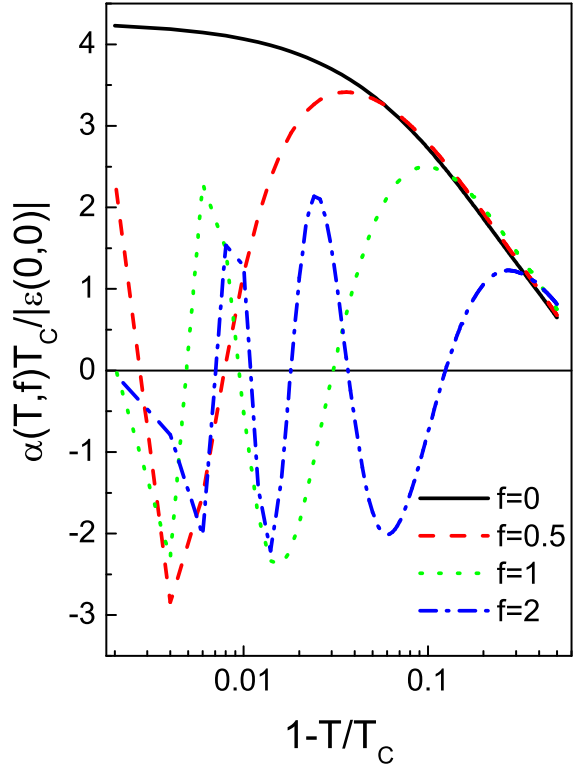}
\caption{\label{fig:2} (Color online) Temperature dependence of
flux driven normalized TEC $\alpha (T,f)T_C/|\epsilon (0,0)|$ in a
single small contact for different values of the frustration
parameter $f=H/H_0(0)$ (for the same set of parameters as in
Fig.1) according to Eqs.(1)-(13).}
\end{figure*}
From the very structure of Eqs.(1)-(9) it is obvious that TEC of a
single contact will exhibit {\it field} oscillations imposed by
the Fraunhofer dependence of the critical current $I_C$. Much less
obvious is its temperature dependence. Indeed, Fig.~\ref{fig:1}
presents the temperature behavior of the contact area strain field
$\epsilon (T,f)$ (with $t(0)/\xi = 1$, $\xi /\lambda _L(0)=0.02$
and $\beta =0.1$) for different values of the frustration
parameter $f=H/H_0(0)$. Notice characteristic flux driven
temperature oscillations near $T_C$ which are better seen on a
semi-log plot shown in Fig.~\ref{fig:2} which depicts the
dependence of the properly normalized field-induced TEC $\alpha
(T,f)$ as a function of $1-T/T_C$ for the same set of parameters.

\section{III. Thermal expansion in the presence of screening currents}

To answer an important question how the neglected in the previous
Section screening effects will affect the above-predicted
oscillating behavior of the field-induced TEC, let us consider a
more realistic situation with a junction embedded into an array
(rather than an isolated contact) which is realized in
artificially prepared arrays using photolithographic technique
(that nowadays allow for controlled manipulations of the junctions
parameters~\cite{array-artificial}). Besides, this is also a good
approximation for a granular superconductor (if we consider it as
a network of superconducting islands connected with each other via
Josephson links~\cite{orlando}). Our goal is to model and simulate
the elastic response of such systems to an effective stress
$\sigma$ (described in the previous Section for an isolated
contact). For simplicity, we will consider an array with a regular
topology and uniform parameters (such approximation already proved
useful for describing high-quality artificially prepared
structures~\cite{a6}).

\subsection{A. Model equations for a planar square array}

Let us consider a planar square array as shown in
Fig.~\ref{fig:3}. The total current includes the bias current
(flowing through the vertical junctions) and the induced screening
currents (circulating in the plaquette~\cite{nakajima}). This
situation corresponds to the inclusion of screening currents only
into the nearest neighbors, neglecting thus the mutual inductance
terms~\cite{phillips}. Therefore, the equation for the vertical
contacts will read (horizontal and vertical junctions are denoted
by superscripts $h$ and $v$, respectively):
\begin{figure*}
\includegraphics[width=6.5cm,angle=90]{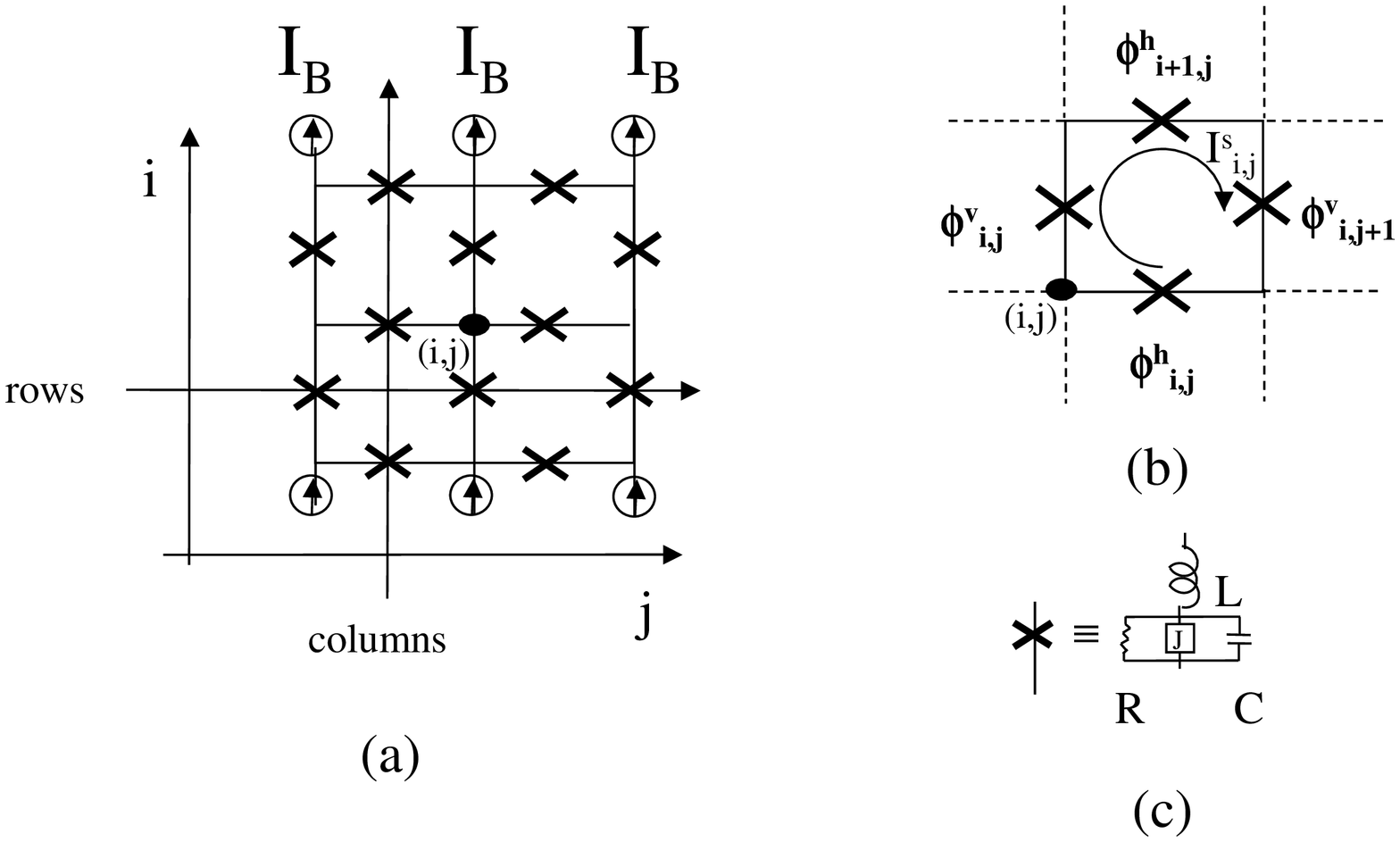} \vskip 1cm
\caption{\label{fig:3} Sketch of an array. The junctions
perpendicular (parallel) to the bias are called horizontal
(vertical). (a) The node $(i,j)$ is shown as a circle in the left
bottom corner of a plaquette; (b) a single plaquette (the
elementary unit of the circuit) along with the circulating
current; and (c) the lumped elements circuit for a small
junction.}
\end{figure*}
\begin{equation}
\frac{\hbar C}{2e}\frac{d^2\phi_{i,j}^v}{dt^2} + \frac{\hbar}{2eR}
\frac{d\phi_{i,j}^v}{dt} +
 I_c \sin \phi_{i,j}^v = \Delta I^s_{i,j} + I_b,
\label{currentcons-nostress}
\end{equation}
where $\Delta I^s_{i,j}=I^s_{i,j}-I^s_{i-1,j}$ and the screening
currents $I^s$ obey the fluxoid conservation condition:
\begin{equation}
-\phi^v_{i,j}+\phi^v_{i,j+1} - \phi^h_{i,j} + \phi^h_{i+1,j} =
2\pi \frac{\Phi^{ext}}{\Phi_0} - \frac{2\pi L I^s_{i,j}}{\Phi_0}.
\label{fluxoid}
\end{equation}

Recall that the total flux has two components (an external
contribution and the contribution due to the screening currents in
the closed loop) and it is equal to the sum of the phase
differences describing the array. It is important to underline
that the external flux in Eq.(\ref{fluxoid}), $\eta = 2\pi
\Phi^{ext}/\Phi_ 0$, is related to the frustration of the whole
array (i.e., this is the flux across the void of the
network~\cite{fernando,grimaldi}), and it should be distinguished
from the previously introduced applied magnetic field $H$ across
the junction barrier which is related to the frustration of a
single contact $f=2\pi Hdw/\Phi _0$ and which only modulates the
critical current $I_C(T,H,\sigma )$ of a single junction while
inducing a negligible flux into the void area of the array.

\subsection{B. Stress induced effects}

For simplicity, in what follows we will consider the elastic
effects due to a uniform (homogeneous) stress imposed on the
array. With regard to the geometry of the array, the deformation
of the loop is the dominant effect with its radius $a$ deforming
as follows:
\begin{equation}
a(\sigma)=a_0(1-\chi \sigma/\sigma_0).
\end{equation}
As a result, the self-inductance of the loop $L(a)=\mu_0 a F(a)$
(with $F(a)$ being a geometry dependent factor) will change
accordingly:
\begin{equation}
L(a)=L_0(1-\chi_g \sigma/\sigma_0).
\label{induct-dep}
\end{equation}
The relationship between the coefficients $\chi$ and $\chi_g$ is
given by
\begin{equation}
\chi_g = \left( 1+a_0 B_g \right)\chi
\label{chirelation}
\end{equation}
where $B_g=\frac{1}{F(a)}\left(\frac{dF}{da}\right)_{a_0}$.

It is also reasonable to assume that in addition to the critical
current, the external stress will modify the resistance of the
contact:
\begin{equation}
R(\sigma ) =\frac{\pi \Delta(0)}{2 e I_C(\sigma )}=R_0 e^{-\chi
\sigma/\sigma_0} \label{resistance}
\end{equation}
as well as capacitance (due to the change in the distance between
the superconductors):

\begin{equation}
C(\sigma )=\frac{C_0}{1-\chi \sigma/\sigma_0}\simeq C_0 (1+\chi\sigma/\sigma_0).
\label{capacitance}
\end{equation}
To simplify the treatment of the dynamic equations of the array,
it is convenient to introduce the standard normalization
parameters such as the Josephson frequency:

\begin{equation}
\omega_J = \sqrt{\frac{2\pi I_C(0)}{C_0 \Phi_0}}, \label{omegaj}
\end{equation}
the analog of the SQUID parameter:

\begin{equation}
\beta_L = \frac{2\pi I_C(0)L_0}{\Phi_0},
\label{squidpar}
\end{equation}
and the dissipation parameter:

\begin{equation}
\beta_C = \frac{2\pi I_C(0)C_0R_0^2}{\Phi_0}.
\label{betac}
\end{equation}
Combining Eqs.(\ref{currentcons-nostress}) and (\ref{fluxoid})
with the stress-induced effects described by Eqs.
(\ref{resistance}) and (\ref{capacitance}) and using the
normalization parameters given by
Eqs.(\ref{omegaj})-(\ref{betac}), we can rewrite the equations for
an array in a rather compact form. Namely, the equations for
vertical junctions read:
\begin{eqnarray}
\frac{1}{1-\chi \sigma/\sigma_0} \ddot{\phi}_{i,j}^v + \frac{e^{-\chi
\sigma/\sigma_0}}{\sqrt{\beta_C}} \dot{\phi}_{i,j}^v +
e^{\chi \sigma/\sigma_0} \sin \phi_{i,j}^v  =  \hspace{6cm} \nonumber \\
\frac{1}{\beta_L \left( 1 -\chi_g \sigma/\sigma_0 \right) }
 \left[ \phi^v_{i,j-1} - 2\phi^v_{i,j}  +  \phi^v_{i,j+1} + \phi^h_{i,j} - \phi^h_{i-1,j} +
\phi^h_{i+1,j-1} - \phi^h_{i,j-1}  \right] + \gamma_b.
\label{arr-eq-vert}
\end{eqnarray}
Here an overdot denotes the time derivative with respect to the
normalized time (inverse Josephson frequency), and the bias
current is normalized to the critical current without stress,
$\gamma_b = I_b/I_C(0)$.

The equations for the horizontal junctions will have the same
structure safe for the explicit bias related terms:
\begin{eqnarray}
\frac{1}{1-\chi \sigma/\sigma_0} \ddot{\phi}_{i,j}^h + \frac{e^{-\chi
\sigma/\sigma_0}}{\sqrt{\beta_C}} \dot{\phi}_{i,j}^h +
e^{\chi \sigma/\sigma_0} \sin \phi_{i,j}^h =  \hspace{6cm} \nonumber \\
\frac{1}{\beta_L \left( 1 -\chi_g \sigma/\sigma_0 \right) } \left[
\phi^h_{i,j-1} -2\phi^h_{i,j}+\phi^h_{i,j+1} + \phi^v_{i,j} -
\phi^v_{i-1,j} + \phi^v_{i+1,j-1} - \phi^v_{i,j-1} \right].
\label{arr-eq-hor}
\end{eqnarray}
Finally, Eqs.(\ref{arr-eq-vert}) and (\ref{arr-eq-hor}) should be
complemented with the appropriate boundary
conditions~\cite{binder} which will include the normalized
contribution of the external flux through the plaquette area $\eta
= 2\pi \frac{\Phi^{ext}}{\Phi_0}$.

It is interesting to notice that Eqs.(\ref{arr-eq-vert}) and
(\ref{arr-eq-hor}) will coincide with their stress-free
counterparts if we introduce the stress-dependent re-normalization
of the parameters:

\begin{equation}
\tilde{\omega}_J = \omega_J e^{\chi \sigma /2\sigma_0}, \label{omegajren}
\end{equation}

\begin{equation}
\tilde{ \beta}_C = \beta_C e^{- 3 \chi \sigma/\sigma_0}, \label{betacren}
\end{equation}

\begin{equation}
\tilde{\beta}_L = \beta_L (1-\chi_g \sigma/\sigma_0) e^{\chi \sigma/\sigma_0 },
\label{squidparren}
\end{equation}

\begin{equation}
\tilde{\eta} = \eta (1- 2 \chi \sigma/\sigma_0 ), \label{etaren}
\end{equation}

\begin{equation}
\tilde{\gamma}_b = \gamma_b e^{-\chi \sigma/\sigma_0 }.
 \label{gammaren}
\end{equation}

\subsection{C. Numerical results and discussion}

Turning to the discussion of the obtained numerical simulation
results, it should be stressed that the main problem in dealing
with an array is that the total current through the junction
should be retrieved by solving self-consistently the array
equations in the presence of screening currents.
Recall~\cite{orlando} that the Josephson energy of a single
junction for an arbitrary current $I$ through the contact reads:

\begin{equation}
E_J(T,f,I)=E_J(T,f,I_C)\left[1-\sqrt{1-\left(\frac{I}{I_C}\right)^2}
\right]. \label{Josenergybias}
\end{equation}

\noindent The important consequence of Eq.(\ref{Josenergybias}) is
that if no current flows in the array's junction, such junction
will not contribute to the TEC (simply because a junction
disconnected from the current generator will not contribute to the
energy of the system).

\noindent  Below we sketch the main steps of the numerical
procedure used to simulate the stress-induced effects in the
array:

\begin{itemize}
\item[1)] a bias point $I_b$ is selected for the whole array;
\item[2)] the parameters of the array (screening, Josephson frequency,
dissipation, etc) are selected and modified according to the
intensity of the applied stress $\sigma$;
\item[3)] the array equations are simulated to retrieve the static configuration of
the phase differences for the parameters selected in step $2$;
\item[4)] the total current flowing through the individual junctions is
retrieved as:

\begin{equation}
I^{v,h}_{i,j} = I_C\sin \phi^{v,h}_{i,j}; \label{totalcurrent}
\end{equation}

\item[5)] the energy dependence upon stress is numerically estimated using
the value of the total current $I^{v,h}_{i,j}$ (which is not
necessarily identical for all junctions) found in step $4$ via
Eq.(\ref{totalcurrent});
\item[6)] the array energy $E_J^A$ is obtained by summing up the contributions of
all junctions with the above-found phase differences
$\phi^{v,h}_{i,j}$;
\item[7)] the stress-modified screening currents $I^s_{i,j}(T,H,\sigma )$
are computed by means of Eq.(\ref{fluxoid}) and inserted into the
magnetic energy of the array $E_M^A = \frac{1}{2L}
\Sigma_{i,j}(I^s_{i,j})^2$;
\item[8)] the resulting strain field and TE coefficient of the array are
computed using numerical derivatives
based on the finite differences:

\begin{equation}
\epsilon^A \simeq \frac{1}{V} \left[
\frac{\Delta\left(E_M^A+E_J^A\right)}{\Delta
\sigma}\right]_{\Delta\sigma \rightarrow 0},
 \label{estimstrain}
\end{equation}

\begin{equation}
\alpha (T,H) \simeq \frac{ \Delta\epsilon^A}{ \Delta T}.
\label{estimalpha}
\end{equation}
\end{itemize}
\begin{figure*}
\includegraphics[width=8.5cm]{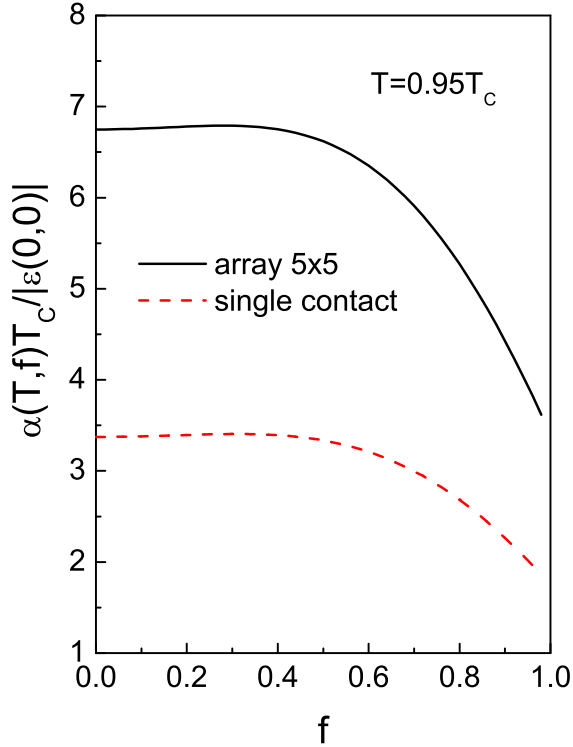}
\caption{\label{fig:4} (Color online) Numerical simulation results
for an array $5 \times 5$ (black solid line) and a small single
contact (red dashed line). The dependence of the normalized TEC
$\alpha (T,f)T_C/|\epsilon (0,0)|$ on the frustration parameter
$f$ (applied magnetic field $H$ across the barrier) for the
reduced temperature $T/T_C=0.95$. The parameters used for the
simulations: $\eta =0$, $\beta = 0.1$, $t(0)/\xi = 1$,
$\xi/\lambda_L=0.02$, $\beta_L=10$, $\gamma_b=0.95$, and
$\chi_g=\chi=0.01$.}
\end{figure*}
The numerical simulation results show that the overall behavior of
the strain field and TE coefficient in the array is qualitatively
similar to the behavior of the single contact. In Fig.~\ref{fig:4}
we have simulated the behavior of both the small junction and the
array as a function of the field across the barrier of the
individual junctions in the presence of bias and screening
currents. As is seen, the dependence of $\alpha (T,f)$ is very
weak up to $f\simeq 0.5$, showing a strong decrease of about $50
\%$ when the frustration approaches $f=1$.
\begin{figure*}
\includegraphics[width=7.2cm]{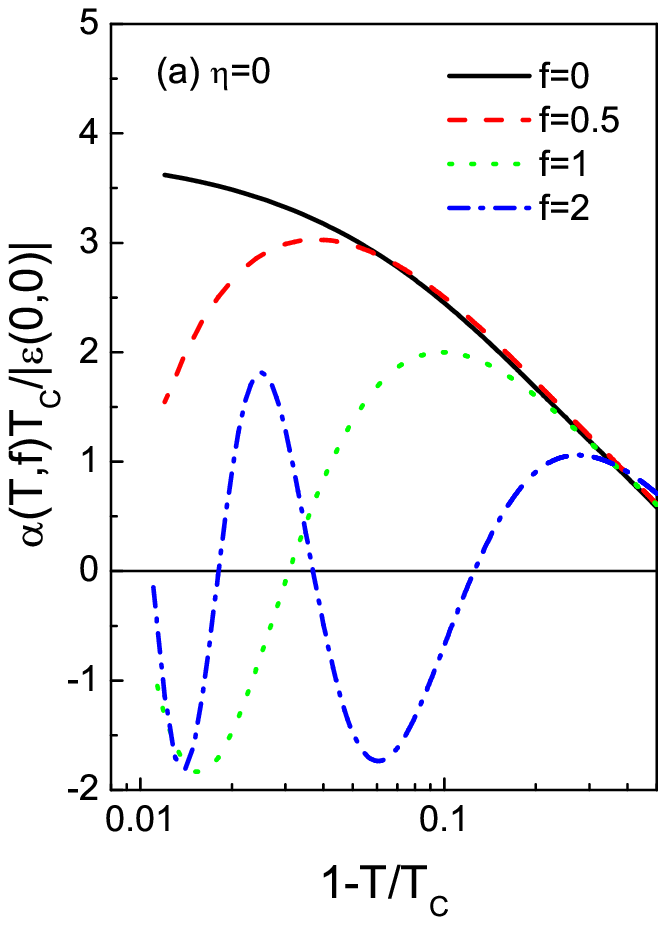}
\includegraphics[width=7.5cm]{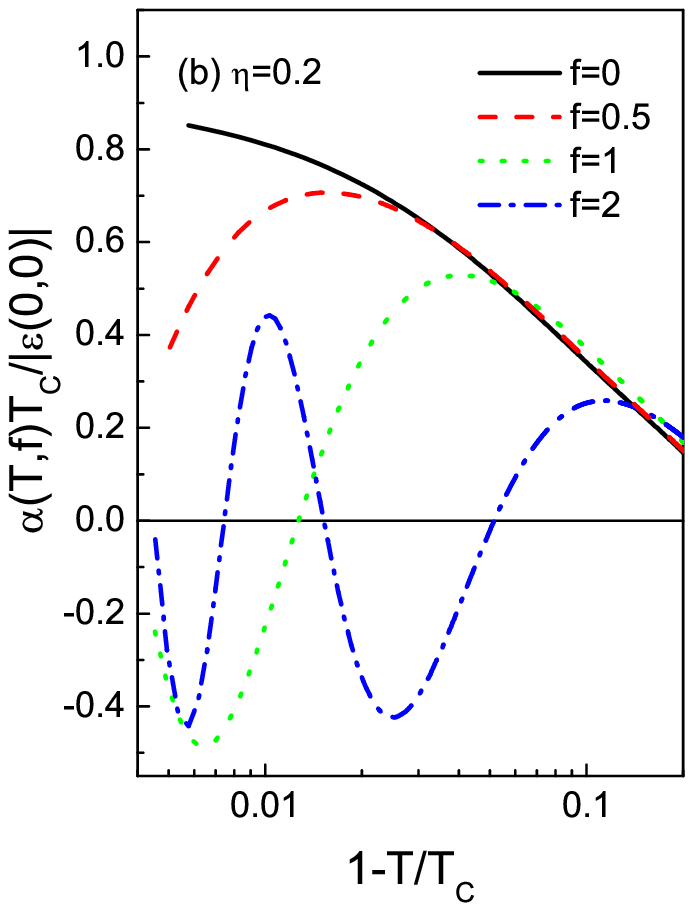}
\includegraphics[width=7.8cm]{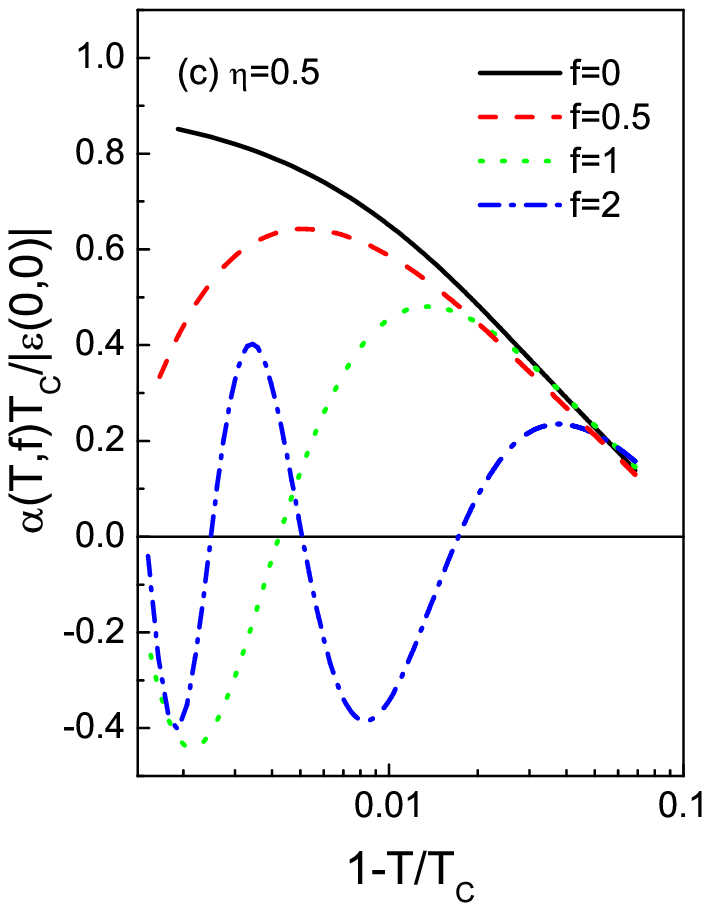}
\includegraphics[width=7.5cm]{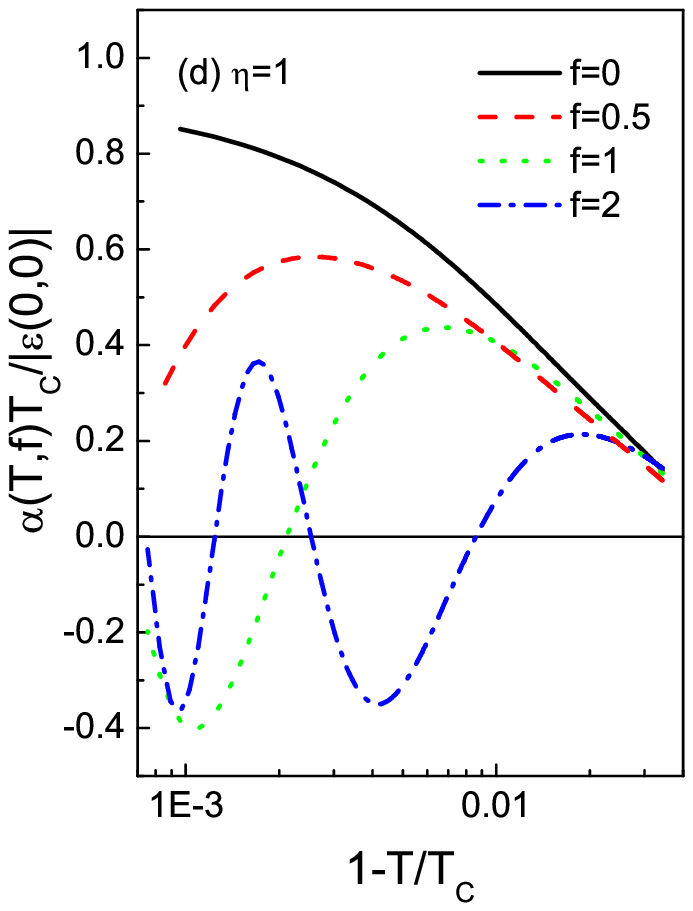}
\caption{\label{fig:5} (Color online) Numerical simulation results
for an array $5 \times 5$. The influence of the flux across the
void of the network $\eta$ frustrating the whole array on the
temperature  dependence of the normalized TEC $\alpha
(T,f)T_C/|\epsilon (0,0)|$ for different values of the barrier
field $f$ frustrating a single junction for $\gamma_b=0.5$ and the
rest of parameters same as in Fig.4.}
\end{figure*}
A much more profound change is obtained by varying the temperature
for the fixed value of applied magnetic field. Fig.~\ref{fig:5}
depicts the temperature behavior of $\alpha (T,f)$ (on semi-log
scale) for different field configurations which include barrier
field $f$ frustrating a single junction and the flux across the
void of the network $\eta$ frustrating the whole array. First of
all, comparing Fig.~\ref{fig:5}(a) and Fig.~\ref{fig:2} we notice
that, due to substantial modulation of the Josephson critical
current $I_C(T,H)$ given by Eq.(\ref{Josenergy}), the barrier
field $f$ has similar effects on the TE coefficient of both the
array and the single contact including temperature oscillations.
However, finite screening effects in the array result in the
appearance of oscillations at higher values of the frustration $f$
(in comparison with a single contact). On the other hand,
Fig.~\ref{fig:5}(b-d) represent the influence of the external
field across the void $\eta$ on the evolution of $\alpha (T,f)$
(recall that in the absence  of stress this field produces a
well-defined magnetic pattern~\cite{fernando,grimaldi,binder}). As
is seen, in comparison with a field-free configuration (shown in
Fig.~\ref{fig:5}(a)), the presence of external field $\eta$
substantially  reduces the magnitude of the TE coefficient of the
array. Besides, with $\eta$ increasing, the onset of temperature
oscillations markedly shifts closer to $T_C$.

\section{IV. Conclusion}

We have investigated the influence of a homogeneous mechanical
stress on a small single Josephson junction and on a plaquette
(array of $5\times 5$ junctions). We have shown how the
stress-induced modulation of the parameters describing the
junctions (as well as the connecting circuits) produces such an
interesting phenomenon as a thermal expansion (TE) in a single
contact and two-dimensional array (plaquette). We also studied the
variation of the TE coefficient with an external magnetic field
and temperature. In particular, near $T_C$ (due to some tremendous
increase of the effective "sandwich" thickness of the contact) the
field-induced TE coefficient of a small junction exhibits clear
{\it temperature} oscillations scaled with the number of flux
quanta crossing the contact area. Our numerical simulations
revealed that these oscillations may actually still survive in an
array if the applied field is strong enough to compensate for
finite screening induced self-field effects. And finally, it is
important to emphasize that our analysis refers to regular arrays
with square geometry (similar to already existing artificially
prepared arrays~\cite{a6,fernando}). However, we can argue that
the predicted here effects should manifest themselves also in
granular superconductors which exhibit quite pronounced stress
dependent behavior upon mechanical loading~\cite{a9,2}.

\begin{acknowledgments}
We are thankful to the Referee for very useful comments which
helped improve the presentation of this paper and better
understand the obtained here results. SS and FMAM gratefully
acknowledge financial support from the Brazilian agency FAPESP
(Projeto 2006/51897-7). GF and GR wish to acknowledge financial
support from the CNISM-INFM-CNR Progetto Supercalcolo 2006 and by
ESF in the framework of the network-programme: Arrays of Quantum
Dots and Josephson Junctions.
\end{acknowledgments}

\end{document}